# Monolithic Magneto-Optical Mach–Zehnder Isolator Using Laser-Annealed Iron Garnet on a Silicon Waveguide

**Tomoya Sugita[1], Reona Motoji[1], Yuki Yoshihara[1], Dan Maeda[1], Hiroki Yamamoto[1], Hibiki Miyashita[2], Kazushi Ishiyama[2], Member, IEEE, and Taichi Goto[2], Member, IEEE**

[1]Kyocera Corporation, 3-7-1 Minatomirai, Nishi, Yokohama, Kanagawa, 220-0021, Japan.

[2]Research Institute of Electrical Communication, Tohoku University, 2-1-1, Katahira, Aoba, Sendai, Miyagi, 980-8577, Japan.

Corresponding author: Taichi Goto (e-mail: taichi.goto.a6@tohoku.ac.jp).

This work was supported in part by JSPS KAKENHI under Grant 23K17758, Grant 23K26133 and Grant 26K22455, in part by NEDO under Grant JPNP20004, in part by the Amada Foundation, and in part by the Inamori Foundation.

**ABSTRACT** Stable silicon photonic circuits require monolithically integrated optical isolators based on magneto-optical garnet. However, crystallizing the garnet by conventional furnace annealing exposes the entire chip to high temperature and degrades the silicon waveguides and the metal electrodes. Here we avoid this degradation by using local laser annealing in vacuum to crystallize cerium-substituted yttrium iron garnet (Ce:YIG), deposited by ion beam sputtering without a seed layer, directly within a silicon-based Mach–Zehnder interferometer. A 915 nm beam heats only the garnet region confined in micrometer-scale trenches, leaving the surrounding circuit and electrodes intact. The device achieves an isolation ratio of 13.6 dB at a wavelength of 1540 nm, corresponding to a Faraday rotation of 0.092°/μm, with an insertion loss of 20.4 dB and a propagation loss of 9.5 dB. Transmission electron microscopy reveals the crystallized Ce:YIG and a ~10 nm boundary region at the interface with the Si waveguide. These results demonstrate that thermally sensitive silicon photonic devices and magneto-optical thin films requiring high-temperature processing can be integrated by a high-throughput technique compatible with mass production.

**INDEX TERMS** Annealing, Garnet films, Lasers, Magnetooptic devices, Monolithic integration, Sputtering

## I. INTRODUCTION

Silicon-based photonic integrated circuits (PICs) are now widely deployed, particularly in data center optical interconnects, because of their low power consumption and high communication speed [1]. Co-packaged optics (CPO) [2], integrating optical engines with electronic ICs in a single package, is a representative example. For stable device operation, monolithic magneto-optical (MO) isolators are required, but their integration has been challenging because of the high-temperature treatment of MO garnet films possessing a large MO effect and low optical loss [3]. After the first demonstration of a monolithic MO isolator based on a silicon ring resonator using polycrystalline cerium-substituted yttrium iron garnet (Ce:YIG) with a YIG seed layer [4], many improved MO isolators have been reported alongside the development of novel fabrication techniques [3, 5, 6]. For instance, Du et al. [7] formed the MO garnet before the Si waveguide to prevent waveguide degradation, demonstrating an excellent isolator. Shoji et al. [8] bonded a single-crystalline Ce:YIG film, grown epitaxially on a garnet substrate, to a Si waveguide, achieving isolator performance superior to that of polycrystalline films. Goto et al. found vacuum annealing of Ce:YIG to enable seed-layer-free fabrication [9], demonstrating an isolator with a simple process and structure compatible with Si waveguides [10]. Recently, separate groups led by Zhang [11], Sugita [12], and Kim [13] have reported isolators combining Mach–Zehnder interferometers with polycrystalline garnet, achieving high performance through optimized deposition conditions while keeping a simpler structure suited to mass production. However, all these approaches expose the entire sample chip to temperatures exceeding ~600 °C, whether by heating during deposition [14-17] or by furnace annealing afterward [18], so degradation of the Si waveguide is unavoidable and remains a major source of performance loss [19]. Furthermore, the aluminum and copper electrodes used in CPO pose a further obstacle to integration at such temperatures. In contrast to such global heating, laser annealing has long been used to crystallize garnet films [20, 21] to locally modify their magnetic properties [22, 23]. Therefore, in this work, we use vacuum laser annealing [24] for local heating and crystallization of seed-layer-free Ce:YIG, realizing a monolithic Mach–Zehnder isolator. In contrast to the global heating of earlier approaches, this local annealing confines the heat to the garnet region [25-27] and preserves both the Si waveguide and the metal electrodes.

## II. DEVICE FABRICATION

### A. PHOTONIC CIRCUIT WITH CE:YIG THROUGH TRENCH

We fabricated a waveguide MO isolator based on an asymmetric Mach–Zehnder interferometer (AMZI). Light split into two equal halves by a multimode interference (MMI) coupler acquires a phase difference in the two MZI arms of different lengths and recombines at a second MMI coupler. The two arms carry MO material of equal length that produces a nonreciprocal phase shift. The arm-length difference producing the reciprocal phase shift was 15 µm. The waveguide cross section was 180 nm × 550 nm, and the circuit was built on a silicon-on-insulator (SOI) platform by a conventional silicon photonics process (APPENDIX A). The MO material, Ce:YIG, was integrated into the Si waveguide through a trench window. A trench was opened in the cladding to expose the top of the Si waveguide, and Ce:YIG was then deposited to fill the window and overlap the waveguide, forming the MO section. The trench formation and the Ce:YIG deposition are described in APPENDIX B and APPENDIX C, respectively. Sample chips with MO trench lengths of 50, 100, and 200 µm were prepared to study the trench-length dependence. Fig. 1(c) shows the resulting AMZI circuit, and the overall process flow is described in APPENDIX A.

### B. LOCAL LASER ANNEALING

The deposited Ce:YIG was crystallized by local laser annealing in vacuum. Fig. 1(a) shows the laser annealing setup, with the equipment listed in APPENDIX A. The beam from a diode laser at a wavelength of 915 ± 15 nm, with a maximum output of 120 W, was shaped into a 700 µm square and irradiated onto the chip through an anti-reflection-coated window. The chamber was pumped down to below 80 Pa and held there during annealing. The 700 µm square beam covered only the AMZI and its trench, and the electrode regions lay outside the irradiated area, so the surrounding circuit and the electrodes were not heated. A thermometer aligned coaxially with the beam monitored the substrate temperature, and the laser output was regulated to hold the substrate at 750 °C.
Fig. 1(b) shows the annealed chip, and Fig. 1(c) is a magnified view of the region marked in Fig. 1(b). The irradiated region of ~700 µm square turned yellow-green, indicating optical properties different from the as-deposited region. After annealing, the chip was diced to expose the waveguide facets for evaluation.

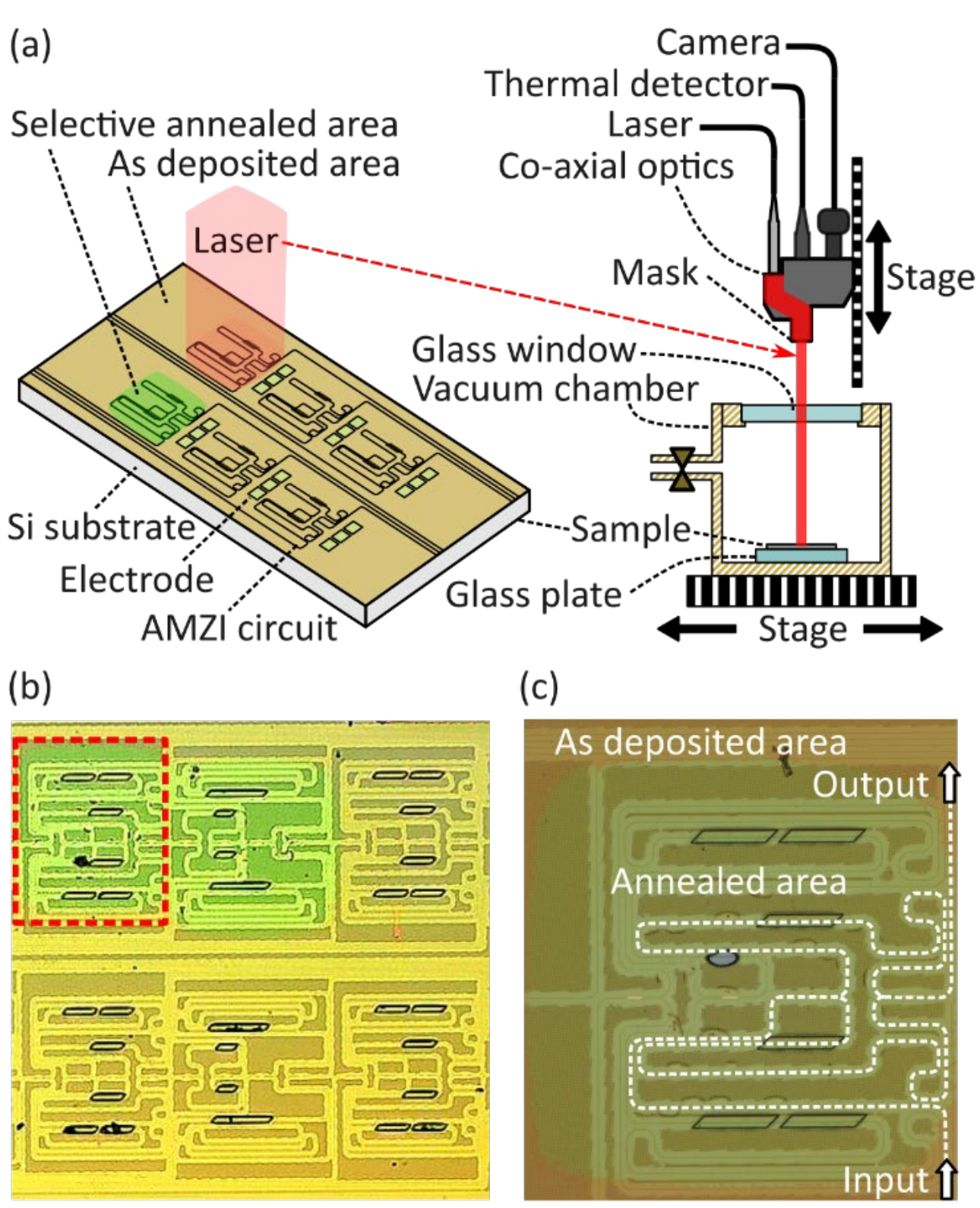

FIGURE 1. (a) Schematic of the laser spot annealing system. (b) Microscope image of the selectively annealed chip. (c) Enlarged microscope image of the selectively annealed area (marked with the red dotted line in Fig. 1(b)). The black parallelograms are MO trenches, and an AMZI circuit is highlighted with a white dotted line.

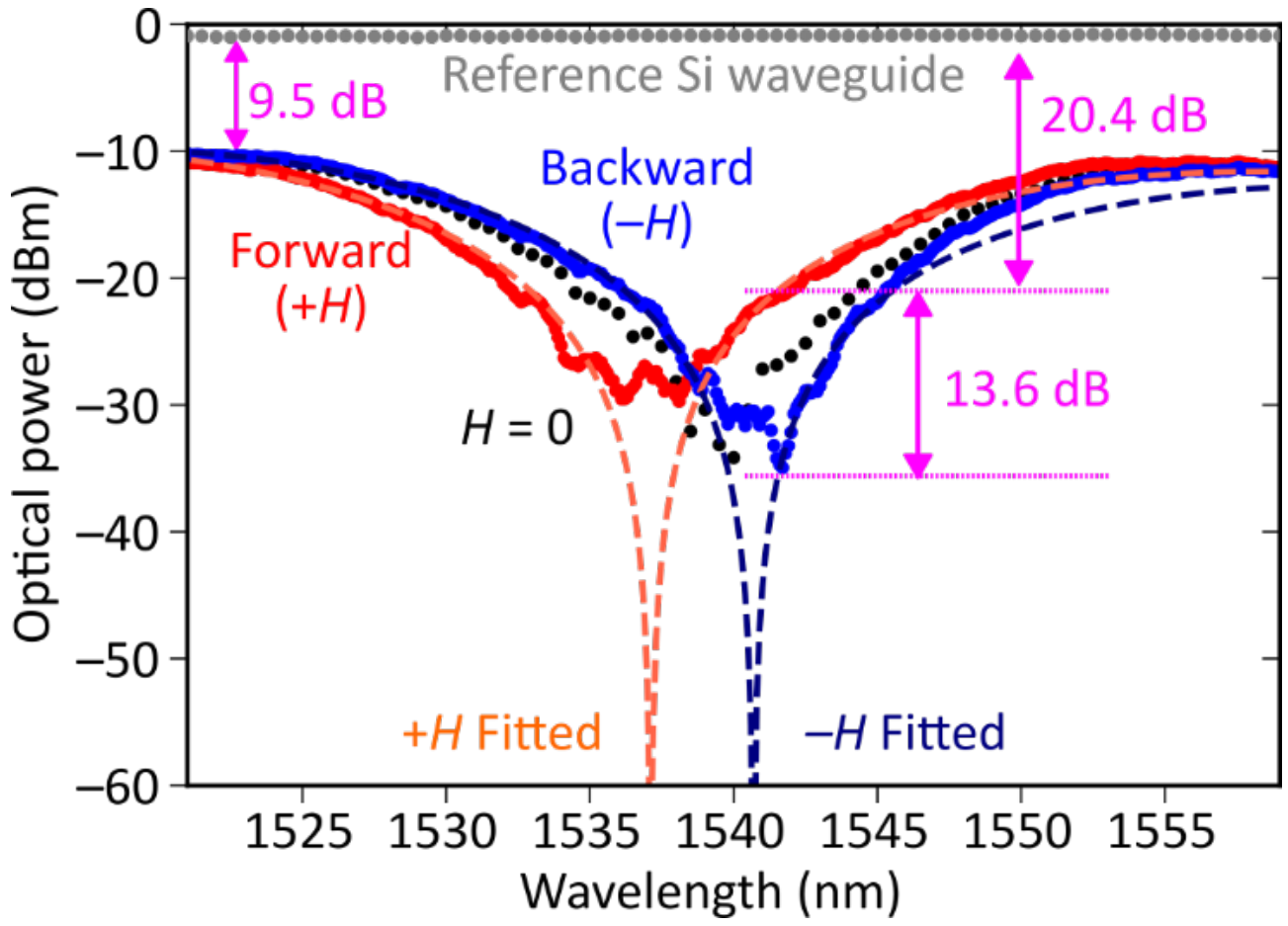

FIGURE 2. Transmission spectra of the MO isolator with a trench length of 100 µm for the forward (red) and backward (blue) directions, with the fitted spectra shown as dashed lines. The gray and black curves are the reference Si waveguide and the no-field result, respectively.

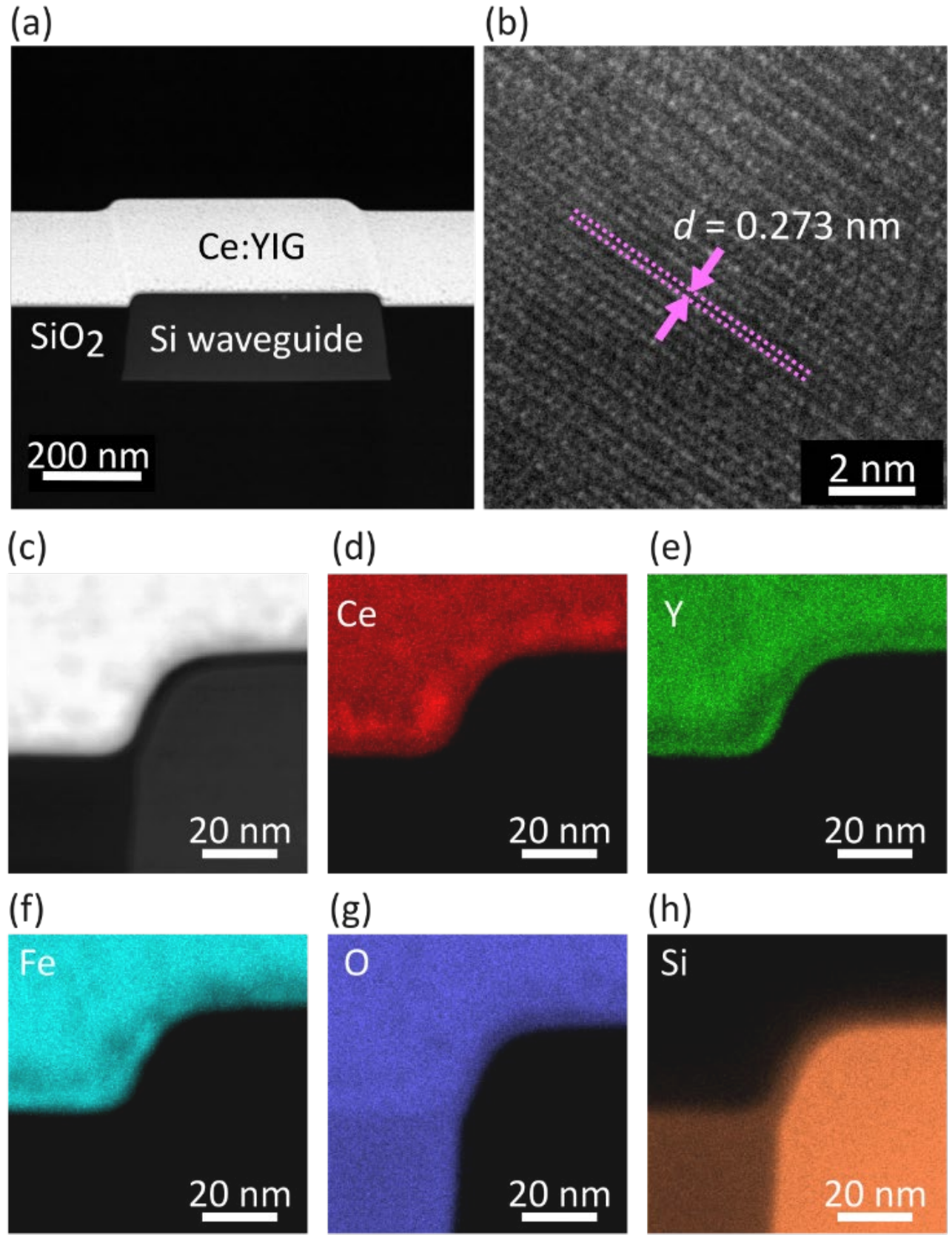

**FIGURE 3. (a) Cross-sectional TEM image of the trench and the Si waveguide, (b) high-resolution TEM image of the Ce:YIG layer over the Si waveguide, (c) cross-sectional bright-field TEM image enlarged to the Ce:YIG/Si-waveguide interface, and (d)–(h) EDS maps of Ce, Y, Fe, O, and Si.**

## III. MO ISOLATOR CHARACTERIZATION

The transmission of the fabricated sample chips was measured by coupling lensed fibers to the input and output, using a tunable laser in the transverse magnetic (TM) polarization and sweeping the wavelength from 1520 to 1610 nm. The sample chip was held at 25 °C. An external field above the saturation field of Ce:YIG was applied in the in-plane direction with SmCo magnets, and the backward response was obtained by reversing the magnet direction with the sample chip fixed, avoiding alignment errors from swapping the fiber. The full setup and procedure are given in APPENDIX D.

## IV. RESULTS AND DISCUSSION

### A. MO ISOLATION DEMONSTRATION

Fig. 2 shows the transmission spectra of the fabricated MO isolator with a trench length of 100 µm for the forward and backward directions. (Backward propagation corresponds to the reflected return light toward the source.) The isolation ratio was 13.6 dB at a wavelength of 1540 nm, taken from the forward–backward difference, and the insertion loss was 20.4 dB. The transmission of the device with no applied magnetic field and that of the reference Si waveguide are also plotted, showing the propagation loss of 9.5 dB. The Faraday rotation (FR) was 0.092°/µm, derived from the forward–backward peak separation of 4.5 nm, the free spectral range (FSR) of 44 nm for the AMZI without MO loading, and the trench length of 100 µm. (The derivation method of the FR value from these quantities is described in Ref. [12].) This FR value is larger than the previously reported 0.01°/µm for laser-annealed Ce:YIG [24], owing to differences in the annealed area and the film quality. In this sample, the Ce:YIG region needed for the FR evaluation was 100 µm × 20 µm, whereas the laser spot was 700 µm square, large enough to anneal almost the entire region uniformly. However, in the previous work, the evaluation region was 5 mm in diameter and the laser spot was only 60 µm, so the region was not fully crystallized and remained non-uniform, indicating the importance of laser intensity uniformity in laser annealing. The obtained FR value of 0.092°/µm was within the range of reported values for Ce:YIG prepared by other methods, 0.021 [11], 0.057 [28], 0.060 [29], and 0.113 [30]°/µm. The origin of the optical loss was analyzed, and most of the loss arises from the additional trench fabrication. The losses attributed to the trench process, the formation of the AMZI, and the Ce:YIG loading were estimated as 66%, 7%, and 27%, respectively (APPENDIX E).

### B. STRUCTURAL ANALYSIS

Fig. 3(a) is a cross-sectional transmission electron microscopy (TEM) image of the MO section in the trench. The Ce:YIG layer was 195 nm thick, and the waveguide was 180 nm high and 550 nm wide. Fig. 3(b) is a high-resolution TEM image of the MO region. The visible atomic layers and the (420) garnet plane, identified from the lattice spacing $d$ = 0.273 nm, close to the reported value of 0.278 nm [18], confirm the crystallization of the Ce:YIG. Fig. 3(c) is a magnified bright-field TEM image of the Ce:YIG/Si-waveguide interface, and (d)–(h) are the energy-dispersive X-ray spectroscopy (EDS) maps of Ce, Y, Fe, O, and Si. The composition ratio Ce:Y:Fe:O was 1:2:5:(12 − δ), where δ is an unmeasured oxygen deficiency, close to the reported value [18]. Within the 5 nm interface region, Ce segregated whereas Y and Fe were deficient. A $SiO_2$ interface layer ~5 nm thick covering the waveguide appears in the O and Si maps and is attributed to a native oxide. Therefore, laser annealing formed a homogeneous garnet on the Si waveguide, apart from the ~10 nm thick Ce/Fe/Y/$SiO_2$ boundary region.

### C. DISCUSSION

The loss analysis (APPENDIX E) attributed ~70% of the propagation loss to the trench process. Notably, dividing the MO region into micrometer-scale trenches prevented the cracking seen when laser annealing areas larger than 1 mm square [24]. The TEM results revealed good-quality Ce:YIG crystals on the silicon photonic circuit, but a ~10 nm interface layer between the Ce:YIG and the Si waveguide was observed.

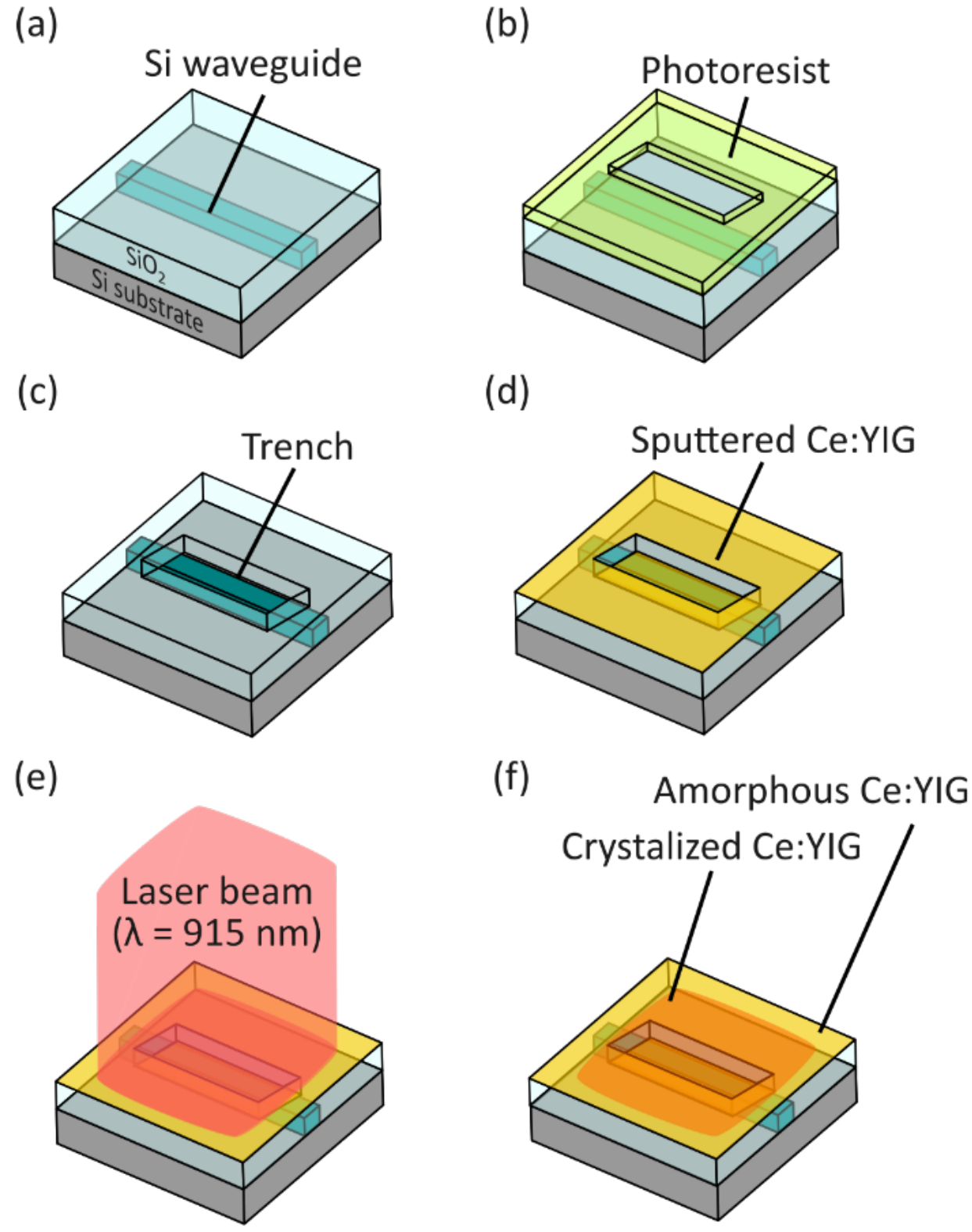

**FIGURE 4.** **Schematic of the fabrication process of the laser-annealed MO optical isolator. (a) silicon PIC with a $SiO_2$ cladding layer, (b) photoresist coating, (c) trench etching by ICP-RIE, (d) Ce:YIG deposition by RF-IBS, (e) selective laser annealing, and (f) completed process.**

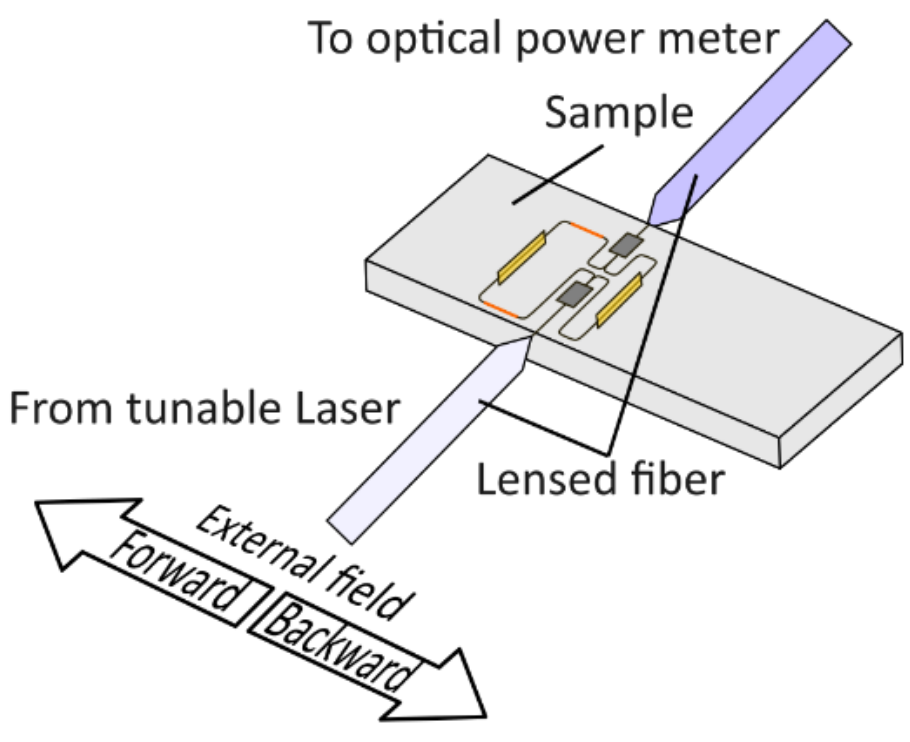

**FIGURE 5.** **A schematic image of the optical isolation measurement system. An external field is switched by swapping directions of SmCo magnet.**

Reducing the propagation loss and enhancing the MO performance therefore requires improving the Si processing, by exposing the waveguide top more precisely through a higher etching selectivity and by suppressing damage to the Si waveguide, and modifying the Ce:YIG, by reducing the $Ce^{4+}$ in the garnet through the oxygen partial pressure and by optimizing the laser annealing conditions.

## V. CONCLUSION

We demonstrated a monolithic MO Mach–Zehnder isolator by using local laser annealing in vacuum to crystallize seed-layer-free Ce:YIG within a silicon PIC. The 915 nm beam heated only the garnet region, preserving the Si waveguides and metal electrodes unlike the global heating of furnace annealing. The device showed an isolation ratio of 13.6 dB at 1540 nm, an insertion loss of 20.4 dB, and a propagation loss of 9.5 dB, with TEM confirming the crystallized Ce:YIG and the interface with the Si waveguide. The process demonstrated here accelerates the integration of MO devices into PICs and contributes to the development of more advanced CPO.

## APPENDIX A
## OVERALL PROCESS FLOW

The silicon PIC was fabricated on an SOI wafer by a commercial silicon photonics process, with a straight reference waveguide ~4.9 mm long on the same chip. The chip was diced from the wafer into 20 mm squares before post-processing. Fig. 4 shows the overall process: (a) the silicon PIC covered with a 5 μm $SiO_2$ cladding layer, prepared by a similar process described in [12], (b) photoresist coating, (c) trench etching by inductively coupled plasma reactive ion etching (ICP-RIE, APPENDIX B), (d) Ce:YIG deposition by radio-frequency ion beam sputtering (RF-IBS, APPENDIX C), (e) selective laser annealing, and (f) the completed process. The optical isolation measurement and the insertion loss analysis are described in APPENDIX D and APPENDIX E, respectively. The laser annealing used a diode laser (AMADA Weldtech, ML-5120A) and an emission unit (AMADA Weldtech, FOCH-30B). The chamber was mounted on an XY stage and the emission unit on a Z stage, with the irradiation position and focus set by a coaxial visible laser and a CCD camera. The chamber was evacuated below 80 Pa by a rotary pump, and the substrate temperature was monitored by a thermal sensor coaxial with the laser.

## APPENDIX B
## TRENCH FABRICATION PROCESS

A G-line positive photoresist (Tokyo Ohka, OFPR800LB-200cp) was spin-coated at 2000 rpm for 20 s (Actes, ASC-4000). The resist thickness was 3.5 μm. A parallelogram trench of 100 μm × 20 μm was exposed with a maskless aligner (Heidelberg Instruments, MLA150), with the short axis tilted by 45° to suppress optical loss from unwanted facet reflection. Samples with various trench lengths of 50 μm, 100 μm, and 200 μm were also fabricated. The resist was developed by immersion in 2.38% tetramethylammonium hydroxide (TMAH) for 3 min and rinsed with deionized water. ICP-RIE with a CE-300I tool (Ulvac) and $CF_4$ and $CHF_3$ gases then exposed the top surface of the Si waveguide and formed the trench.

## APPENDIX C
## MO GARNET DEPOSITION CONDITIONS

Ce:YIG was deposited over the entire chip by RF-IBS using an RM17-0010 system (RMTec). The base pressure was $5 \times 10^{-5}$ Pa and the working pressure was $2.0 \times 10^{-2}$ Pa. The substrate was rotated at 4.3 rpm during deposition. As process gases, Ar at 10 and 14 sccm were introduced to the ion gun and the neutralizer, respectively. $O_2$ at 8 sccm was introduced to the chamber. The film thickness was ~200 nm at a deposition rate of 4.15 nm/min, using a sintered 4-inch $Ce_1Y_{2.5}Fe_5O_x$ target.

## APPENDIX D
## OPTICAL ISOLATION MEASUREMENT SETUP

Fig. 5 shows the measurement setup. A tunable laser (Agilent, 8164B) was the light source, with an input power of 10 dBm. A polarizer preset to the TM detection direction and an infrared camera were placed at the output, and the input fiber was rotated about the optical axis to align the input light with the TM polarization. The sample chip was fixed on a vacuum-chuck stage and held at 25 °C with a Peltier controller (OHM, OCE-TCR24300WL). The output fiber was connected to a power meter (Optotest, OP735) through a 3 dB coupler, and the other branch was monitored with a power meter (ThorLabs, BRM100). The wavelength was swept from 1520 to 1610 nm with a resolution of 0.1 nm. SmCo magnets of 30 mm × 16 mm × 5 mm were arranged to sandwich the sample chip and the stage. A magnetostatic simulation predicted a magnetic field of 250 mT at the center of the facing magnets, above the saturation field of Ce:YIG, and a magnetic field sensor (KANETEC, TM-801) confirmed 220 mT at the sample chip position. A field applied in the left direction relative to the propagation direction was defined as positive, and the opposite as negative. The backward transmission was measured by keeping the sample chip orientation fixed and reversing the magnet direction.

## APPENDIX E
## LOSS ANALYSIS

Fig. 6 shows the transmission of the reference Si waveguide, the reference AMZI, and the MO AMZI without an external field. The no-field MO AMZI result is identical to the black plot in Fig. 2(a). The reference AMZI was fabricated on the same chip and has no additional trench processing. Fig. 6 inset shows the propagation characteristics versus trench length for a Si waveguide with an added trench, for trench lengths of 0, 20, 50, 100, 300, and 500 µm, with a linear-fit slope of $-2.4 \times 10^{-3}$ dB/µm. The reference Si waveguide had a propagation loss of 0.91 dB. Relative to this reference, the total propagation loss of 9.5 dB for the MO AMZI with a 100 µm trench decomposes into formation of the AMZI (0.64 dB, 7%), the trench process (6.3 dB, 66%, comprising the trench-facet loss of 6.1 dB and the trench-bottom loss of ~0.24 dB), and the Ce:YIG loading (2.6 dB, 27%). These contributions are indicated by the markers (i)–(iv) in Fig. 6, corresponding to the reference Si waveguide (i), the formation of the AMZI (ii), the trench process (iii), and the Ce:YIG loading (iv).

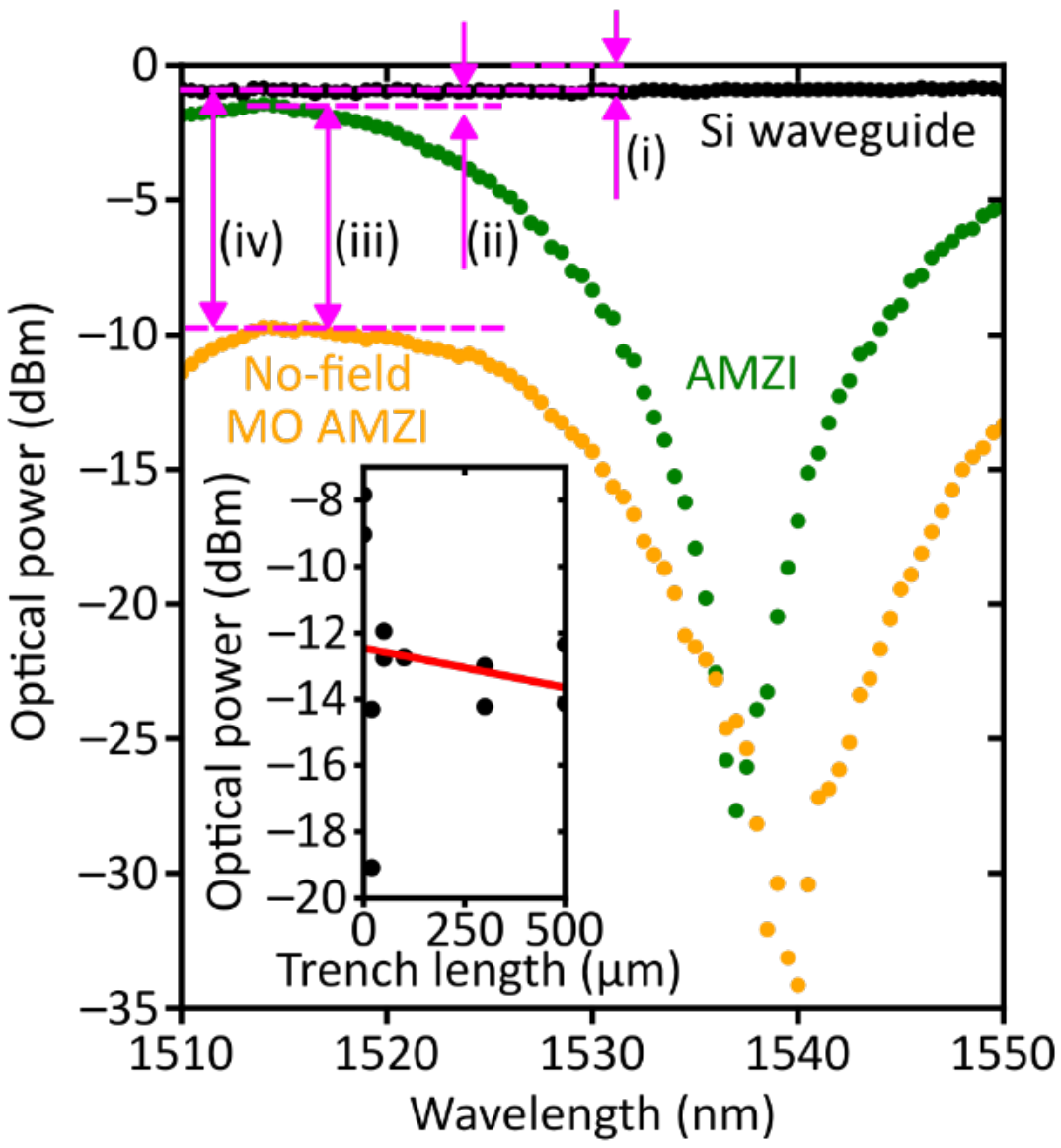

**FIGURE 6.** Optical transmission spectra of the reference Si waveguide, the reference AMZI, and the no-field MO AMZI. Inset shows trench-length dependence of the optical propagation characteristics for a Si waveguide with a trench. The red line is a linear fit.

## ACKNOWLEDGMENT

T.G. acknowledges the fellowship of TI-FRIS. The authors also acknowledge the support of TU-RIEC-CRP, TU-FTC, TU-LNS, TU-CINTS, and TU-GIMRT.

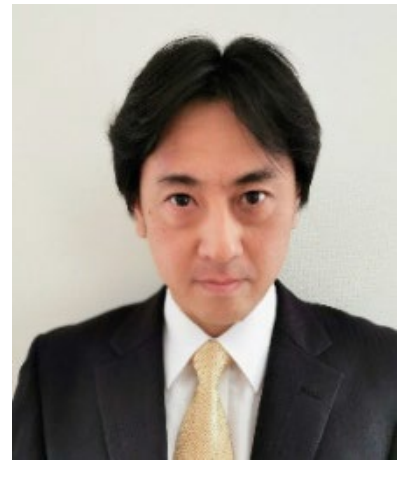

**TOMOYA SUGITA** received B.S. and M.S. degrees in mechano-aerospace engineering and energy sciences from Tokyo Institute of Technology (now Institute of Science Tokyo), Meguro, Japan, in 2000 and 2002, respectively. Since 2002, he has been with Kyocera Corporation. His research interests include the development of optical systems, such as lens optics and photonic devices. He is currently a Manager of Optical Research Section, the Advanced Technology Research Institute, Kyocera Corporation, Yokohama, Japan.

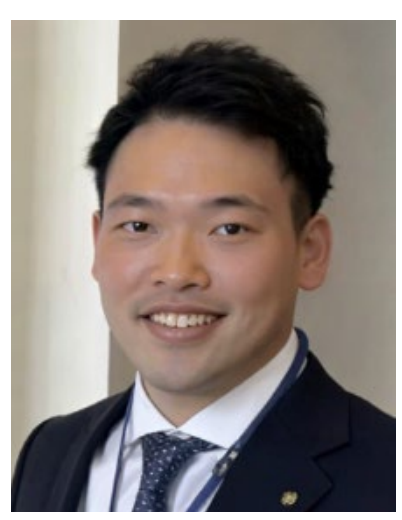

**REONA MOTOJI** received the M.S. degree in materials science from University of Tsukuba, Tsukuba, Japan, in 2019. Since 2019, he has been with Kyocera Corporation, where he is currently a member of the Photonics Business Development Division, Yokohama, Japan.

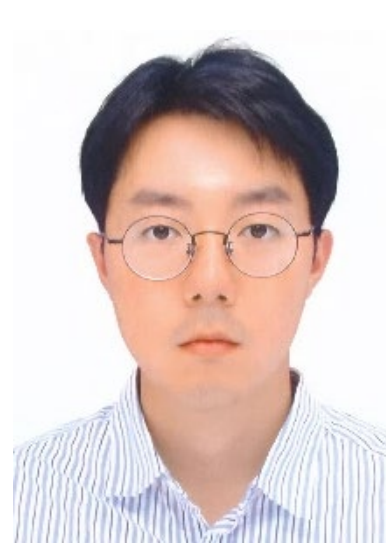

**YUKI YOSHIHARA** received the M.S. degree in electrical and electronic engineering and the Ph.D. degree from the Graduate School of Engineering, Toyohashi University of Technology, Toyohashi, Japan, in 2021 and 2024, respectively. From 2022 to 2024, he was a Research Student at the Research Institute of Electrical Communication, Tohoku University, Sendai, Japan. Since 2024, he has been a member of the Advanced Technology Research Institute, Kyocera Corporation, Yokohama, Japan.

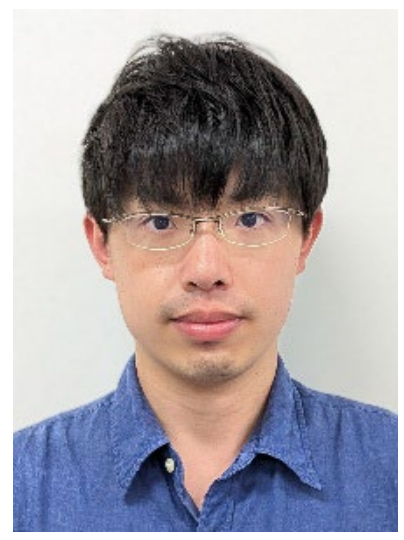

**DAN MAEDA** received the M.S. degree in applied chemistry from the Graduate School of Engineering, The University of Tokyo, in 2021. Since 2021, he has been with Kyocera Corporation, where he is currently a member of the Photonics Business Development Division, Yokohama, Japan.

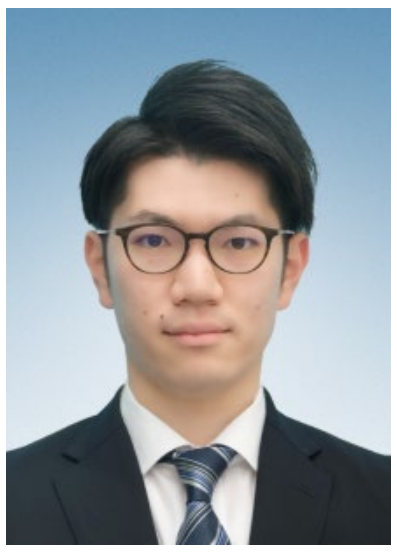

**HIROKI YAMAMOTO** received the M.S. degree in materials engineering from Osaka Prefecture University, Osaka, Japan, in 2023. Since 2023, he has been with Kyocera Corporation, where he is currently a member of the Photonics Business Development Division, Yokohama, Japan.

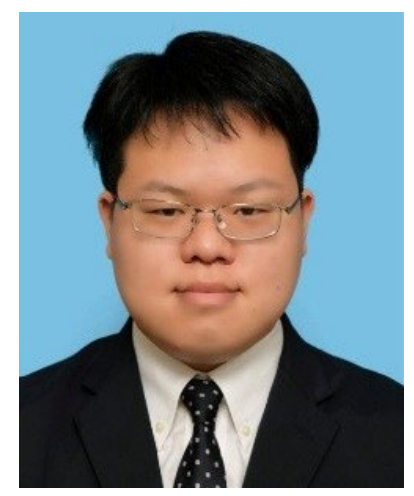

**HIBIKI MIYASHITA** received the M.S. degree in electrical and electronic engineering from Toyohashi University of Technology, Toyohashi, Japan, in 2023, and the Ph.D. degree in electrical engineering from Tohoku University, Sendai, Japan, in 2026.

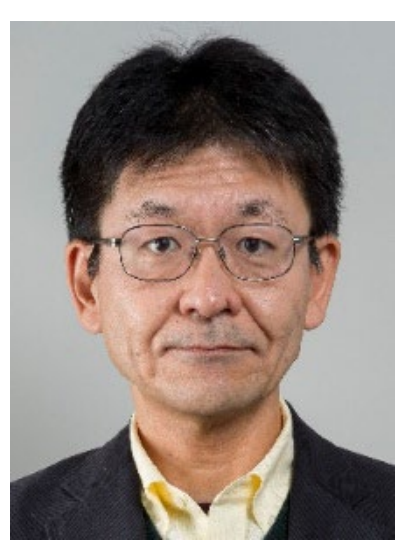

**KAZUSHI ISHIYAMA** (M'93) received the B.S., M.S., and Ph.D. degrees in electrical engineering from Tohoku University, Sendai, Japan, in 1986, 1988, and 1993, respectively. Since 2007, he has been a Professor with the Research Institute of Electrical Communication, Tohoku University, where he has also served as the Director since 2025. His research interests include magnetic devices, such as actuators and sensors, and magnetic materials for these devices. In 2026, he received the Commendation for Science and Technology (Awards for Science and Technology: Research Category) by the Minister of Education, Culture, Sports, Science and Technology.

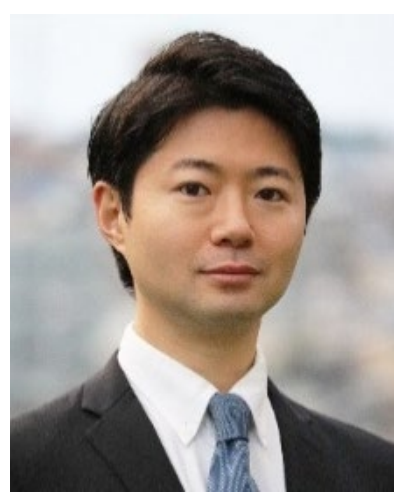

**TAICHI GOTO** (M'08) received the M.S. degree in electrical and electronic engineering and the Ph.D. degree from the Graduate School of Engineering, Toyohashi University of Technology, Japan, in 2009 and 2011, respectively. He was a JSPS Research Fellow (DC1 and PD), including an overseas fellowship, and from 2011 to 2013, he was a Postdoctoral Associate with Massachusetts Institute of Technology (MIT), Cambridge, MA, USA. From 2013 to 2022, he was an Assistant Professor with Toyohashi University of Technology. He was also a JST PRESTO Researcher, from 2015 to 2019, and a Visiting Researcher with MIT, from 2016 to 2020. Since 2022, he has been an Associate Professor with the Research Institute of Electrical Communication, Tohoku University, Sendai, Japan. He also holds the titles of a Tohoku University Distinguished Researcher and a TI-FRIS Fellow at the Frontier Research Institute for Interdisciplinary Sciences. His work on functional microdevices based on magnetic interference was recognized with the Young Scientists' Prize from the Commendation for Science and Technology by the Minister of Education, Culture, Sports, Science and Technology (MEXT), Japan, in 2019. His research interests include spin wave devices, magnetooptical devices, and the development of related magnetic and photonic materials.